\documentclass[pre,twocolumn]{revtex4}

\usepackage{graphicx}
\usepackage{acronym}

\def\v#1{{\mathbf #1}}
\def\x{\v{x}}
\def\TMC{T_\mathrm{MC}}
\def\N{\mathcal{N}}
\def\Z{\mathcal{Z}}
\def\eth{e_\mathrm{th}}

\acrodef{EF}{eigenvector following}
\acrodef{IC}{instantaneous configuration}
\acrodef{INM}{instantaneous normal modes}
\acrodef{IS}{inherent structure}
\acrodef{MCT}{Mode Coupling theory}
\acrodef{MD}{molecular dynamics}
\acrodef{PES}{potential energy surface}
\acrodef{SGM}{square-gradient minimization}
\acrodef{SP}{saddle point}
\acrodef{TMC}[$\TMC$]{the Mode Coupling critical temperature}

\begin{document}

\title{Geometrical properties of the potential energy of the soft-sphere
  binary mixture}

\author{Tom\'as S.\ Grigera}

\thanks{Mailing address: INIFTA, c.c.~16, suc.~4, 1900 La Plata, Argentina}

\affiliation{Instituto de Investigaciones Fisicoqu\'\i{}micas
  Te\'oricas y Aplicadas (INIFTA), Facultad de Ciencias Exactas,
  Universidad Nacional de La Plata, Argentina}

\affiliation{Facultad de Ingenier\'\i{}a, Universidad Nacional de La
  Plata, Argentina}

\affiliation{Consejo Nacional de Investigaciones Cient\'\i{}ficas y
  T\'ecnicas (CONICET), Argentina}

\date{September 12, 2005}

\begin{abstract}

  We report a detailed study of the stationary points (zero-force
  points) of the \ac{PES} of a model structural glassformer. We
  compare stationary points found with two different algorithms
  (eigenvector following and square gradient minimization), and show
  that the mapping between instantaneous configuration and stationary
  points defined by those algorithms is as different as to strongly
  influence the instability index $K$ vs.\ temperature plot, which
  relevance in analyzing the liquid dynamics is thus questioned. On
  the other hand, the plot of $K$ vs.\ energy is much less sensitive
  to the algorithm employed, showing that the energy is the good
  variable to discuss geometric properties of the \ac{PES}. We find
  new evidence of a geometric transition between a minima-dominated
  phase and a saddle-point-dominated one. We analyze the distances
  between instantaneous configurations and stationary points, and find
  that above the glass transition, the system is closer to saddle
  points than to minima.

\end{abstract}

\maketitle

\section{Introduction}

Interest in the geometric properties of the \acf{PES} of liquids as a
mean to understand their dynamics and thermodynamics dates back to the
work of Goldstein \cite{goldstein69} and Stillinger and Weber
\cite{stillinger82}. These works showed that to analyze the low
temperature dynamics of supercooled liquids and glasses it is useful
to map \acp{IC} to the (local) minimum of the \ac{PES} found directly
downhill, called \ac{IS}. This mapping allows to disregard fast
vibrations, focusing on slow, activated, structural relaxations. But
if one aims to describe the dynamic crossover taking place around the
\ac{TMC} \cite{goetze92}, the \ac{IS} mapping is not useful because
barriers between minima are no longer relevant at high temperatures
and the two timescales cease to be well separated. Within the \ac{PES}
approach, two approaches have been proposed. One is to consider whole
superstructures of minimia (called metabasins \cite{doliwa03:hopping,
doliwa03, aldrin03}). Another is to focus on stationary points with
some unstable directions: \acp{SP} \cite{cavagna01}.

The latter approach was motivated by results obtained on the $p$-spin
model (a mean field glass model). In this system, a threshold energy
exists which separates a low-energy, minima-dominated region, from a
high-energy, saddle dominated one \cite{cavagna98, cavagna00}.  The
higher the energy, the larger the number $K$ of unstable directions of
the typical \ac{SP} (this number is called the \emph{order,} or
\emph{instability index} of the \ac{SP}, and is equal to the number of
negative eigenvalues of the Hessian matrix evaluated at the \ac{SP}).
In this model the average index as a function of the energy $K(E)$ can
be computed \cite{cavagna98}.  Furthermore, it can be verified
directly that at high temperatures the stationary point closer to the
typical \ac{IC} is a \ac{SP} with extensive $K$, while it is a minimum
($K=0$) at low temperatures \cite{cavagna01:role}. The dynamic arrest
observed at the dynamic transition is related to the fact that the
system starts getting trapped, or nearly trapped \cite{kurchan96,
franz00}. A similar scenario was then proposed \cite{cavagna01} for
structural glasses: the glass transformation is the consequence of a
geometrical transition.  In the saddle-dominated (high energy) phase,
the system can relax either by jumping an energy barrier or by finding
an unstable direction. In the minima-dominated phase (low energy), the
second mechanism is no longer available; as a consequence relaxation
times soar. It was shown later that the two-step relaxation observed
in the supercooled liquid (and described by \ac{MCT}) can be
qualitatively understood as relaxation in the vicinity of a \ac{SP}
\cite{cavagna03}.

This scenario has been explored in several numerical studies of model
liquids. Some of these works obtained estimations of the $K(E)$ curve
\cite{broderix00, sha01, grigera02}, which have been found compatible
with the existence of a geometric transition (further evidence for the
transition has been found in the context of high frequency vibrations
\cite{grigera03}). Other works have studied instead $K$ as a function
of the temperature $T$ \cite{angelani00, angelani01, sha02, doye02,
wales03}, showing that $K$ decreases dramatically on approaching
\ac{TMC}.  In early studies the view was held that a sharp transition
can be observed as a function of temperature, with $K=0$ for $T<\TMC$
and $K>0$ for $T>\TMC$, but further work has shown \cite{doye02,
fabricius02, doliwa03} that although $K$ decreases very fast (most
likely with an Arrenhius law \cite{doliwa03, berthier03}), it is still
nonzero for $T<\TMC$. This has prompted criticism of the
saddle--minima transition point of view (see e.g.\
ref.~\onlinecite{berthier03}), although a geometric transition,
controled by the energy, is compatible with a smooth $K(T)$ curve (see
sec.~\ref{sec:transition}).

But there is another issue to be discussed when considering $K(T)$
curves. Since the system is never \emph{precisely at a \ac{SP},} to
define a $K(T)$ curve one needs to introduce a mapping between
\acp{IC} and \acp{SP} (in a sense defining a ``basin of attraction''
of a \ac{SP}, and generalizing the \ac{IS} concept).  Additionally, if
one wants to somehow interpret dynamic behavior from such curve, the
mapping should preserve at least some dynamic information.  In
analytical studies (e.g.\ \cite{cavagna01:role, zamponi03,
andronico04}) \acp{IC} are (reasonably) mapped to the nearest \ac{SP}
(using the Euclidean distance or some overlap function).  In contrast,
in numerical work \acp{IC} are mapped to a \ac{SP} through the
algorithm used to find the latter, thus in principle introducing a
dependence on the details of the procedure used to find \acp{SP}
\cite{zamponi03, andronico04} and raising the question of the
dynamical relevance of the mapping.  This is perhaps more worrying
given that at least one popular procedure fails rather often, leaving
some configurations unmapped (see ref.~\onlinecite{doye02} and
discussion below). Also puzzling are some results \cite{doye02,
wales03} that seem to indicate that in the typical distance from an
instantaneous configuration to a saddle or to a minimum is the same,
at variance with the mean field situation.

In this paper we address the issue of the mapping between \acp{IC} and
\acp{SP}, and analyze of distances between \acp{IC} and \acp{SP} and
minima in more detail than has been previously done. Our results show
that the $K$ {\sl vs.}\ $T$ plots are algorithm-dependent, and that,
at least in the soft-sphere model we consider, \acp{IC} at high
temperature are closer to \acp{SP} with $K>0$. The large number of
\acp{SP} collected allows a new analysis which provides new evidence
for the existence of a geometrical transition.

\section{Model and algorithms}

\label{methods}

We have considered the soft sphere binary mixture \cite{bernu87,
barrat90}, which consists of 50\% of particles of type $A$ and 50\% of
type $B$, interacting with a pair potential
$v_{ij}(r)=(\sigma_i+\sigma_j)^{12}/r^{12}$. The radii $\sigma_i$ are
fixed by the conditions $\sigma_B/\sigma_A=1.2$ and $(2\sigma_A)^3 + 2
(\sigma_A+\sigma_B)^3 + (2\sigma_B)^3 = 4$. We have used a system of
$N=70$ particles at unit density and a smooth (cubic polynomial)
long-range cut-off at $\sqrt{3}$ as in ref.~\onlinecite{grigera02}. We
have used swap Monte Carlo \cite{grigera01} to equilibrate the system
at temperatures $T=1$, $0.683$, $0.482$, $0.350$, $0.260$,
$0.220$. For this system \ac{TMC} is about 0.24 \cite{grigera02,
barrat90}.  At each temperature, 40000 equilibrated configurations
were saved and used as starting point for minima and saddle point
searches. Minima were obtained with Nocedal and Liu's LBFGS algorithm
\cite{liu89}, which code can be obtained from the internet
\cite{netlib}. For \ac{SP} searches, two different algorithms were
employed: \ac{SGM} and \ac{EF} (described below), to compare two
different \ac{IC}--\ac{SP} mappings. In all, about $3.2 \cdot 10^5$
\acp{SP} were obtained.

\subsection{Square gradient minimization}

One way of finding \acp{SP} is minimizing the squared modulus of the
gradient,
\begin{equation}
\phi=|\nabla V|^2 = \sum_{i=1}^N \sum_{\alpha=1}^3 \left( {\partial V \over
    \partial x_{i,\alpha} } \right)^2.
\end{equation}
Since $\phi$ is nonnegative, at the absolute minima $\phi=0$, which
implies $\nabla V=0$ (a saddle point). This method is relatively easy
to implement, since good numerical minimization algorithms are
publicly available (we have used LBFGS \cite{liu89, netlib}). The
biggest drawback is that minimization can (and does rather often)
converge to a \emph{local} minimum, which is neither a saddle point,
nor close to one in any reasonable sense \cite{doye03:comment}. 

\subsection{Eigenvector following}

This method has been specifically designed to find stationary points
of the potential energy. Based on an original proposal by Cerjan and
Miller \cite{cerjan81}, it has been substantially improved by others
(see ref.~\onlinecite{wales04:landscape} and references therein). The
problem originally considered \cite{cerjan81} was to find a saddle
point of index 1 starting from a local minimum of $V$. The idea was to
consider the function on a small sphere around the minimum. Using
Lagrange multipliers and a quadratic approximation, one looks for
local minima of the function constrained to the sphere. Close enough
to the initial point (minimum), there is one local minimum of the
constrained function that has higher energy: this is a point along the
path that leads to the sought saddle point, and is taken as the
starting point of the next iteration. Close to the saddle this
criterion no longer applies, so a Newton-Raphson step is taken.

We have used our own implementation of the eigenvector following
method as described by Wales and coworkers \cite{wales93, wales94,
wales96, wales03}. At each iteration a step $\Delta \x$ is proposed,
which in the base that (locally) diagonalizes the Hessian is
\cite{wales93,wales94}
\begin{equation}
  \Delta x_\mu = S_\mu {2 g_\mu \over |h_\mu| \left( 1 + \sqrt{1 + 4 g^2_\mu
        / h^2_\mu } \right) },
\end{equation}
where $h_\mu$ are the eigenvalues of the Hessian and $g_\mu$ are the
components of the gradient in the diagonal base ($\Delta x_\mu$ is set
to 0 for the directions where $h_\mu=0$; i.e.\ uniform displacements).
The sign $S_\mu=\pm 1$ is chosen as explained below. Note that as
$g_\mu\to0$,
\begin{equation}
  \Delta x_\mu = - {g_\mu \over h_\mu} + O(g_\mu^2), \qquad g_\mu\to0
\end{equation}
where the first term is the Newton-Raphson step. A set of trust radii
$\{\delta_\mu\}$ is mantained (one for each direction) \cite{wales96}.
The proposed step is rescaled so that $|\Delta x_\mu| \le \delta_\mu$
for all $\mu$, and then the position is updated. Initially, the
$\delta_\mu$ are set to $0.2$, and at each step are increased
(decreased) by a factor 1.2 according to whether the quantity
$r=(h_e-h_\mu)/h_\mu$ is less (greater) than 1. $h_e$ is an estimation
of the eigenvalue, $h_e=(g_\mu-g'_\mu)/\Delta x'_\mu$, where the prime
means the quantity evaluated at the previous iteration \cite{wales96}.

If $S_\mu=1$, the step increases the energy along the direction $\mu$,
thus the algorithm converges to a maximum along this direction.
Conversely if $S_\mu=-1$ the algrithm converges to a minimum along
$\mu$. Since a saddle point of order $K$ is a maximum along $K$
directions and a minimum along $3N-K$ directions, in princple the
algorithm may be made to converge to a saddle of the desired index by
setting $S_\mu=-1$ for $1\le\mu\le K$, and $S_\mu=1$ for $\mu>K$. In
this work we do not want to fix the index from the start of the
search, so for each starting configurations we run 20 steps with
$S_\mu=-\mathrm{sgn}\, h_\mu$ and only then fix the index to whatever
value it has reached after the first 20 steps \cite{wales03}.

\subsection{Distances}

\label{sec:algo-distance}

The distance we report is the Euclidean distance
\begin{equation}
  d = \sqrt{ \sum_{i,\alpha} (x_{i,\alpha} -
      y_{i,\alpha})^2 },
\end{equation}
minimized over the symmetry operations of the system (i.e.\
translations, the 48 discrete symmetries of the simple cubic lattice,
and particle permutations). Minimization over translations is done
applying Brent's method \cite[\S 10.2]{book:recipes} to a distance
minimized over the discrete symmetries and permutations. This in turn
is found by exhaustive exploration of the discrete symmetries and
using the Hungarian algorithm \cite{kuhn55} (as implemented by
B. Gerkey \cite{hungarian}) to minimize over permutations.

\section{Comparison of SGM and EF mappings}

\subsection{Basins of attraction and success rate}

\begin{figure}
  \includegraphics[width=\columnwidth]{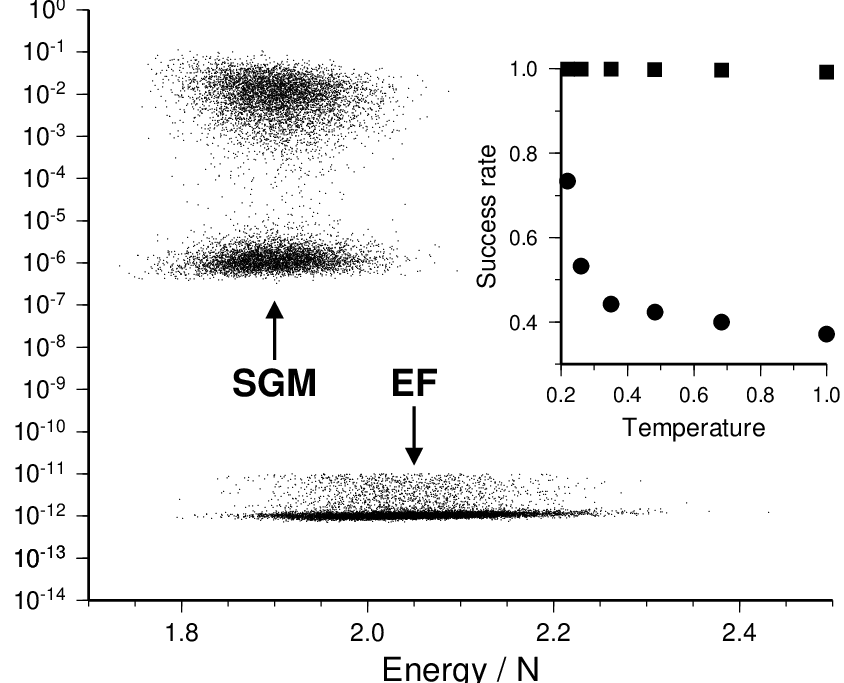}
  \caption{Root-mean-square gradient vs.\ energy of the configuration
    obtained with \ac{SGM} or \ac{EF} starting from 10000 equilibrium
    configurations at $T=0.350$.  Inset: success rate (fraction of
    initial configurations that converge to saddle points) vs.\
    temperature (\ac{SGM}: circles, \ac{EF}: squares).}
  \label{fig:forcerms}
\end{figure}

In Fig.~\ref{fig:forcerms} where we show the rms gradient
$g=\sqrt{\phi /N}$ of the configurations obtained after running
\ac{SGM} and \ac{EF} on the same set of \acp{IC}. \ac{EF} produces
tightly converged saddle points ($g<10^{-11}$), while the
configurations found by \ac{SGM} have rms gradients that cluster
around about $10^{-6}$ and $10^{-2}$. To decide whether these
configurations are saddle points and compute a success rate
(Fig.~\ref{fig:forcerms}, inset) we have used the \ac{SGM}
configurations as starting points for \ac{EF} searches and computed
the distance between the \ac{SGM} and corresponding \ac{EF} converged
configurations. In some cases, after a few (3--5) steps, \ac{EF} found
a saddle very close to the \ac{SGM} connfiguration (distances of the
order of $10^{-5}$ to $10^{-2}$), while in other cases the distance
was $O(1)$ or larger (and the number of steps increased). The first
case clearly corresponds to an absolute minimum of $\phi$, and the
second to a local minimum. The criterion we used to accept the
\ac{SGM} configuration as a true saddle was to require that the
distance between it and the corresponding \ac{EF} saddle be less than
$0.01$. This is approximately equivalent to requiring $g<10^{-4}$ for
the \ac{SGM} configuration.

\begin{figure}
  \includegraphics[width=\columnwidth]{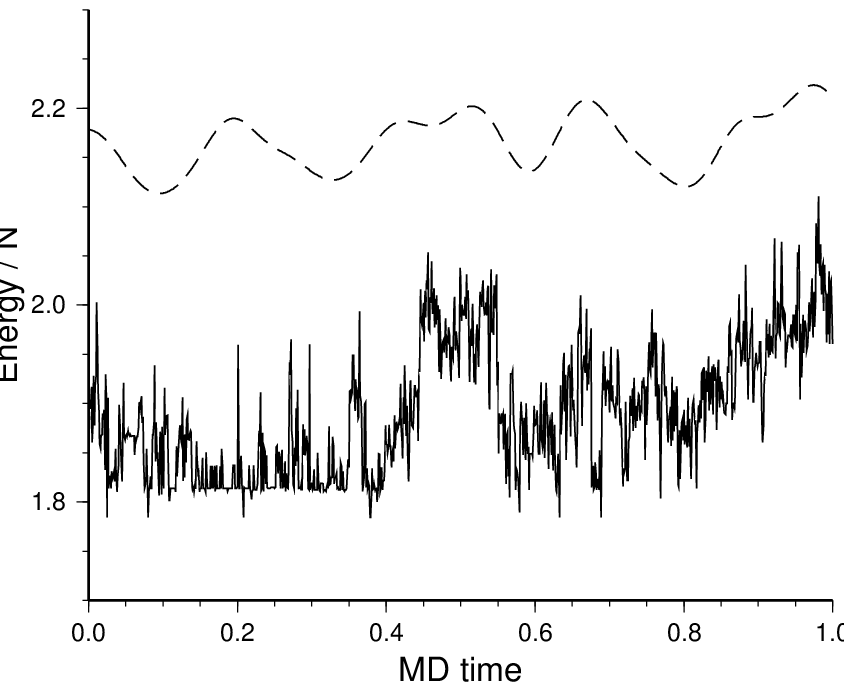}
  \caption{Instantaneous energy along a 1000-step MD run at $T=0.260$
  (dashed-dotted line) and energy of the saddle points found by \ac{EF}
  using the MD configurations as starting points (full line). All
  instantaneous configurations correspond to a single \ac{IS}.}
  \label{fig:stability}
\end{figure}

The high failure rate of the \ac{SGM} algorithm makes it unsuitable to
define basins of attraction of saddle points \cite{doye02, wales03}.
This failure rate is not due to problems with the minimization
algorithm, but to the high number of local minima of the function
$\phi$ \cite{doye02, grigera02, broderix00}. On the other hand the
\ac{EF} algorithm in principle will always converge to a saddle point
(eventual failures being due to numerical problems or implementation
details). However, it should be remembered that although basins of
attraction for saddle points can be defined using \ac{EF}, these are
not necessarily a reasonable generalization of \acp{IS}. It is well
known that iterative non-linear algorithms can lead to multiply
connected or even fractal basins (a case of fractal basins is the
Newton-Raphson algorithm applied to finding the roots of the
polynomial $z^3-1$, see e.g.\ ref.~\onlinecite[\S 9.4]{book:recipes}).
In the case of \ac{EF}, a detailed study on a 3-atom cluster
\cite{wales93} has shown that the basins, though not fractal, are
still complex and multiply connected. Their relevance to liquid
dynamics is thus not to be taken for granted. We have not performed
such detailed analysis here, but in Fig.~\ref{fig:stability} we plot
the instantaneous energy along a short (1000 steps) \ac{MD} run, along
with the energy of the \ac{EF} saddle points found starting from each
\ac{IC}. All these \acp{IC} map to a single \ac{IS} (minimum). The
strong energy fluctuations of the saddles found in this way indicate
that the usefulness of the basins of attraction so defined is probably
rather limited in understanding the liquid dynamics.

\subsection{Saddle index curves}

\begin{figure}
  \includegraphics[width=\columnwidth]{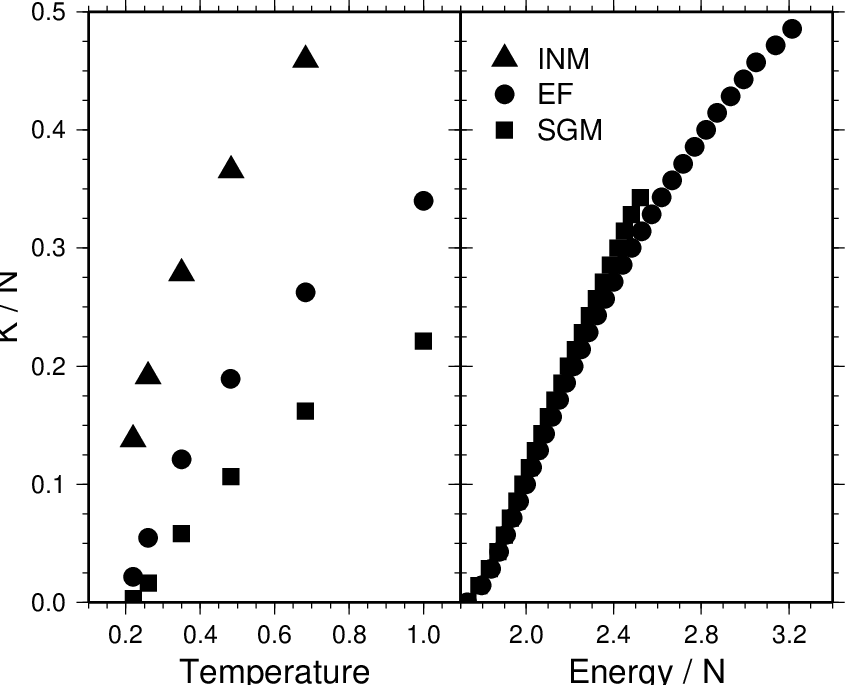}
  \caption{Average instability index vs.\ temperature (left) and
    energy (right) for INM, SGM saddles, and EF saddles. Each point is
    an average over \acp{SP} obtained from 40000 \acp{IC}.}
  \label{fig:kplot}
\end{figure}

Let us first consider the saddle index vs.\ temperature curve.  In
Fig.~\ref{fig:kplot} we plot separately the average value of $K$ for
saddles obtained with \ac{EF} and \ac{SGM}. We also plot $K(T)$
evaluated for \acp{IC} (the eigenvectors of the Hessian evaluated at
an \ac{IC} are usually called \ac{INM}).  The fact that the curves are
algorithm-dependent prevents one from drawing any dynamical conclusion
from them, unless there is some reason to expect that the mapping
between \acp{IC} and \acp{SP} preserves some dynamical information. In
particular, the critical temperature $T_0$ where $K(T_0)=0$ (which
might or might not be greater than zero) will be algorithm-dependent
and thus not of much significance without further evidence of the
dynamical relevance of the algorithm chosen to compute $K(T)$.

\begin{figure}
  \includegraphics[width=\columnwidth]{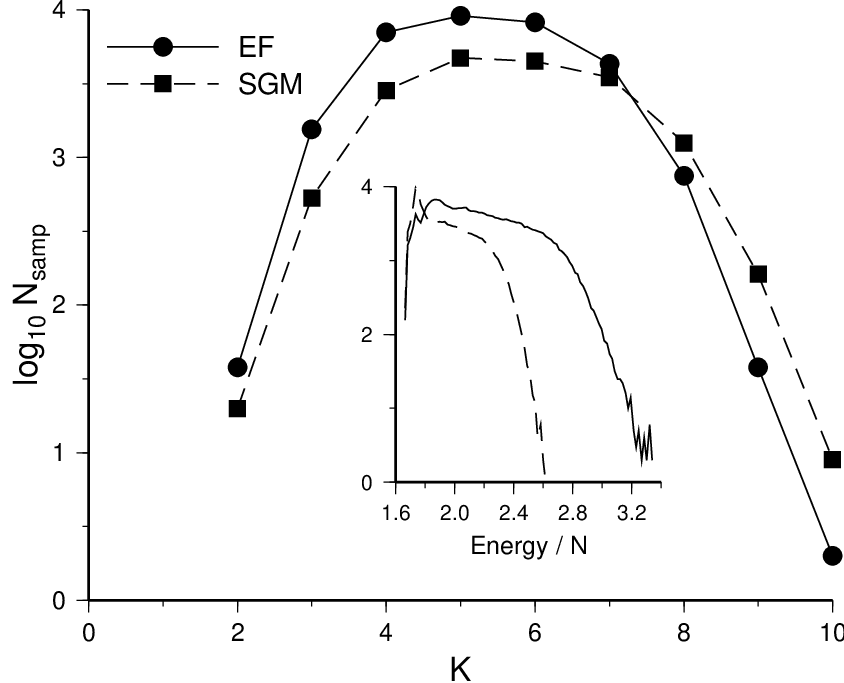}
  \caption{Logarithm of the number of \acp{SP} found vs.\ instability
    index, for the energy band $1.9\le e<2.0$. Inset: Logarithm of the
    number of \acp{SP} vs.\ energy.}
  \label{fig:sampling}
\end{figure}

On the other hand, if we consider $K$ as a funcion of the
\emph{energy} of the saddle point (Fig.~\ref{fig:kplot}, right), the
curves produced from the \acp{SP} collected with \ac{EF} and \ac{SGM}
are essentially coincident in the region of energies where both
algorithms find a significant number of saddles (see inset of
Fig.~\ref{fig:sampling}). The corresponding curve for \acp{INM} (not
shown) is very close to those corresponding to the \acp{SP}, in
contrast to what is found in Lennard-Jones \cite{broderix00}.  In this
purely geometrical plot, the problem of the \ac{IC}-\ac{SP} mapping is
avoided, and issues such as the existence of a \emph{geometrical}
transition can be meaningfully discussed.

Of course, this does not mean one should not worry about possible
biases introduced by the algorithms. Indeed, if one considers the
\acp{SP} within a given energy band, the distribution of the saddle
index is slightly different, with \ac{EF} tending to be slightly
narrower (Fig.~\ref{fig:sampling} shows a representative energy band).
It is also clear that \ac{SGM} tends to find \acp{SP} with lower
energy (inset of Fig.~\ref{fig:sampling}). However, we find that the
maximum of the $\log N_\mathrm{samp}$ vs. $K$ curves are the same for
both algorithms at all energy bands. The significance of this maximum
can be appreciated from the considerations of
sec.~\ref{sec:transition}.

\subsection{Distances}

\label{sect:distance}

Finally we consider the distances between \acp{IC} and the stationary
points (\ac{IS}, \ac{EF} saddles and \ac{SGM} saddles) obtained
starting from the given \ac{IC}. The distances reported here are those
obtained after minimizing over the symmetries of the hamiltonian, as
discussed in sec.~\ref{sec:algo-distance}. We have found that this
distance coincides most of the time with the distance obtained without
applying minimizations when one considers an \ac{IC} and the
stationary point obtained from it, so that the averages we report are
not significantly different from those obtained without minimizing.
Minimization is however important when computing distances between a
configuration and a stationary point obtained from a different
\ac{IC}, as we do below.

\begin{figure}
  \includegraphics[width=\columnwidth]{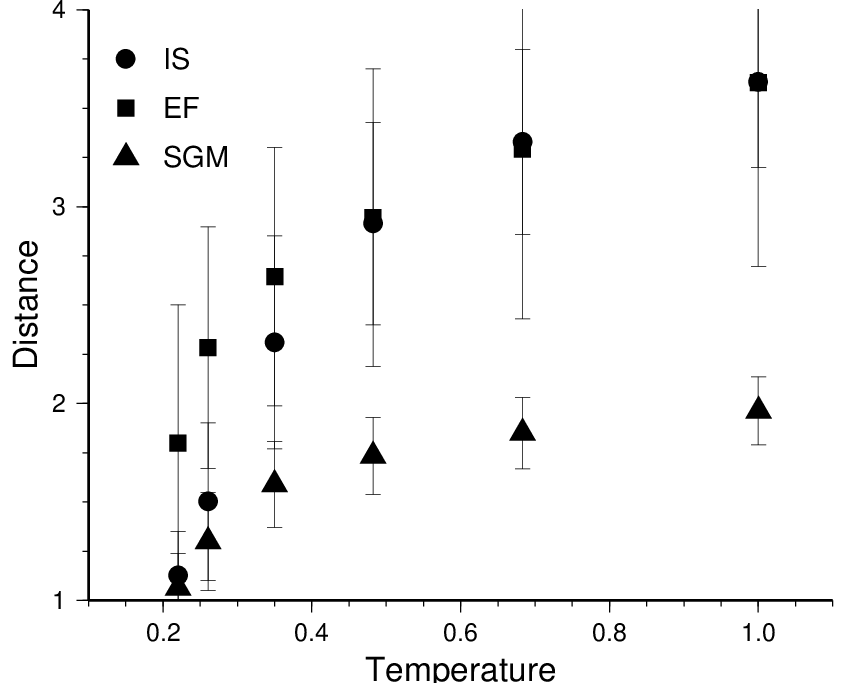}
  \caption{Average distance from instantaneous configurations vs.\
    temperature for minima, EF saddles, and SGM saddles. Error bars
    are estimates of sample standard deviation (not errors on the
    average themselves).}
  \label{fig:dvsT}
\end{figure}

The average distances as a function of temperature are plotted in
Fig.~\ref{fig:dvsT}. Two things are to be noted: first, \ac{SGM}
saddles are always closer than \ac{EF} saddles; second, \acp{IS} are
farther than \ac{SGM} saddles at high temperatures but start to be
found closer as temperature is lowered. The first fact points to the
influence of the algorithm in defining an \ac{IC}--\ac{SP} mapping,
stressing the problems of interpretation of $K(T)$ curves. The second
provides direct evidence that at high temperatures there is a saddle
point to be found closer to the typical \ac{IC} than the corresponding
\ac{IS}.

\begin{figure}
  \includegraphics[width=\columnwidth]{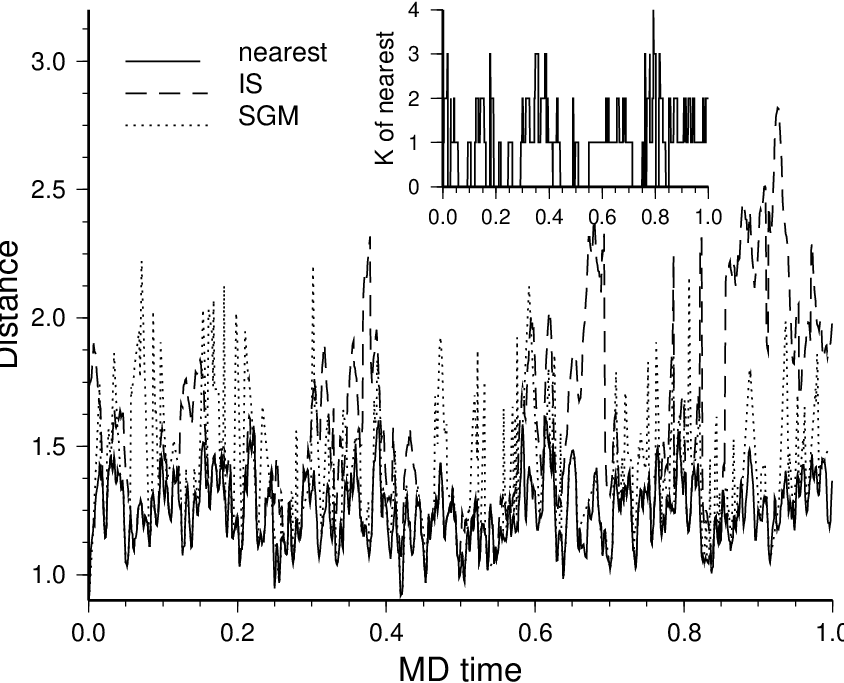}
  \caption{Distance from instantaneous configurations along a short MD
  run at $T=0.26$ to the corresponding IS, SGM saddle and nearest
  \ac{SP} found. Inset: index of nearest \ac{SP}.}
  \label{fig:MDdist}
\end{figure}

To investigate this matter more closely, we look in detail at a short
\ac{MD} trajectory at $T=0.26$, slightly above \ac{TMC}
(Fig.~\ref{fig:MDdist}). For all \acp{IC} in this run, apart from
computing the corresponding \ac{IS} and \ac{SGM} saddle as usual, we
have searched for the nearest \ac{SP} in the pool of all \acp{SP}
found at the corresponding temperature. We find again that the system
is mostly close to a saddle point of order $K\ge1$, as can be also
seen in the inset. This result is natural within the geometrical
transition scenario, but this is, to our knowledge, the first direct
observation of this fact in a liquid.  We furthermore find that the
\ac{SGM} saddle is \emph{not} the closest saddle point. Of course, our
procedure does not guarantee to give the closest saddle to a given
\ac{IC}, but we do find \acp{SP} closer than the \ac{SGM} saddle. We
must note that Wales and Doyle \cite{wales03}, in an analysis of a
Lennard-Jones binary mixture, did not find differences in the mean
distances from \acp{IC} to \acp{SP} or \acp{IS}. It would be
interesting to analyze that system at the single configuration level
(in the spirit of Fig.~\ref{fig:MDdist}).

\section{Saddle--minima transition}

\label{sec:transition}

Consider the number $\N(K,E)$ of saddle points of order $K$ and energy
between $E$ and $E+dE$. For short range interactions, one expects to
be able to divide the system into effectively independent subsystems,
so that $\N(K,E)$ should be exponential in the size of the system
\cite{wales03, shell04}. Then $\log\N$ is extensive in the
thermodynamic limit, and the complexity $\Sigma(k,e)=(1/N)\log\N(K,E)$
is an intensive quantity. The (intensive) average saddle index can
then be written ($e=E/N$)
\begin{eqnarray}
 \langle k(e) \rangle &=& {1\over \Z} \int_0^\infty {K\over N}
 \exp[N\Sigma(K/N,e)] \, dK \\
 &=& {N\over\Z} \int_0^\infty k \exp[N\Sigma(k,e)] \, dk,
\end{eqnarray}
where
\begin{equation}
 \Z = \int_0^\infty \exp[N\Sigma(k,e)] \,dk.
\end{equation}
Using the saddle point method one gets
\begin{equation}
 \langle k(e) \rangle = \hat k(e) + O(1/N),
\end{equation}
where $\hat k(e)$ is the point where $\Sigma(k,e)$ attains a maximum
(with $e$ fixed), i.e.\ the solution of $\partial \Sigma(k,e)/\partial
k=0$. The saddle--minima transition should be understood as happening
at the threshold energy $\eth$, defined as the maximum energy for
which $\hat k(e)=0$. In the thermodynamic limit this implies $\langle
k(e\le \eth) \rangle=0$. In finite systems the average will remain
positive, but the $\langle k(e)\rangle$ curve will show a fast
crossover, remnant of the sharp $N\to\infty$ transition, just as in
other thermodynamic phase transitions. Clearly, the transition does
not mean that there are no saddles for $e<\eth$ (quite the contrary,
there is an exponential number of them), but that they are subdominant
respect to minima: $\N(K>0)/\N(0)\to 0$ exponentially with $N$.

\emph{The control parameter of the (geometric) saddle--minima
transition is the energy \cite{cavagna01, broderix00, sha01,
grigera02, grigera03}.} If one considers $K(T)$ (assuming one defines
a mapping free from the problems discussed above), one will very
likely find $K>0$ below $\TMC$ because relaxation processes, though
slow, will eventually sample saddle points (transition states). A
smooth $K(T)$ curve is thus compatible with a geometric
transition. Within the landscape point of view, one can understand the
sharp dynamic crossover happening in fragile liquids around $\TMC$ as
the signature of a geometric transition \cite{grigera02, grigera03}.

\begin{figure}
  \includegraphics[width=\columnwidth]{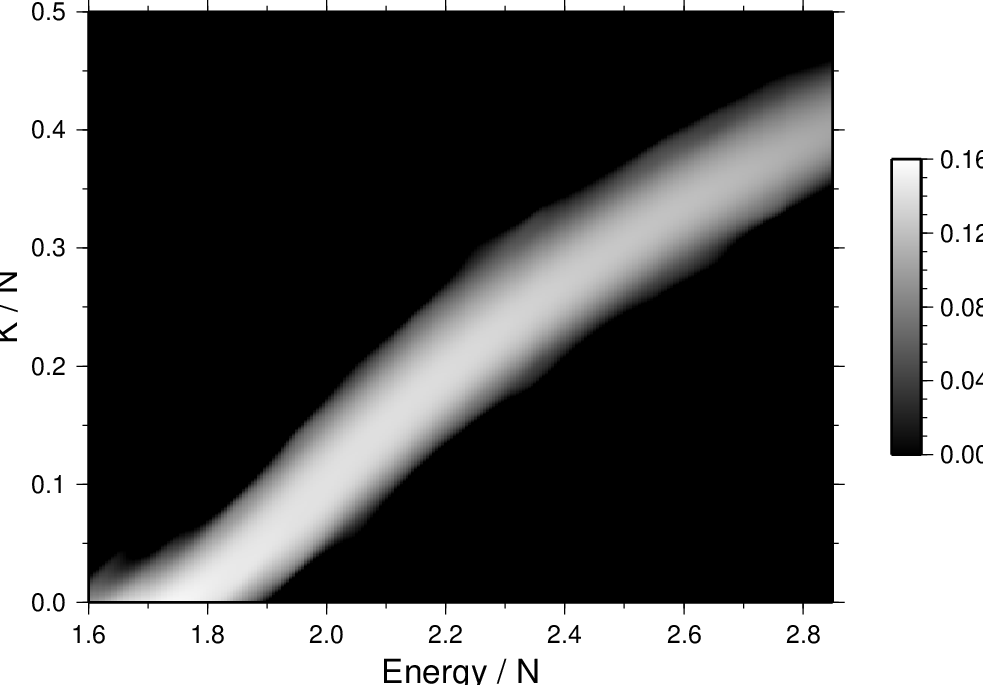}
  \caption{Logarithm of number of saddle points.}
  \label{fig:sigma}
\end{figure}

Given the large number of \acp{SP} obtained for this work (about
$3.2\cdot 10^5$), we can try to obtain a rough estimate of the
qualitative behavior of $\Sigma(k,e)$. The \acp{SP} have been
classified into energy bands of width $0.1$, and for each band a
histogram in $K$ was constructed. The logarithm of the histogram
heights gives an estimate of the shape of the actual $\Sigma$ and is
shown in Fig.~\ref{fig:sigma}. One can clearly see a maximum that goes
to $k=0$ for low values of the energy.  From the present data one
obtains $\eth=1.77 \pm 0.01$, which is compatible with the values
found in refs.~\onlinecite{grigera02} and~\onlinecite{grigera03}.

\section{Conclusions}

\label{conclusions}

We have shown that in numerical studies of the \ac{PES} of liquids,
the algorithm chosen to associate instantaneous configurations and
saddle points can have a significant influence in the analysis of
quantities like the saddle index vs.\ temperature curve. Of the two
algorithms used, we have found that the saddles found with \ac{SGM}
are closer to the instantaneous configurations than those found with
\ac{EF}. Since no such difference was found in
ref.~\onlinecite{wales03}, the present results may reflect a property
of the soft-sphere model, or of the present implementation of the
\ac{EF} algorithm. In any case, the point is that curves such as
$K(T)$ are not meaningful unless the validity of the \ac{IC}--\ac{SP}
mapping with respect to dynamic properties is established. It seems
that neither of these algorithms is useful to define a partition of
phase pase into basins of attraction of a \ac{SP} (in the case of
\ac{SGM} it is rigorously impossible \cite{doye02}).

The natural variable to analyze geometrical properties of the \ac{PES}
is the energy. We have shown that this choice of variable largely
avoids the issue of the \ac{IC}--\ac{SP} mapping (in particular the
$K(E)$ plot is mostly independent of the mapping), though the
algorithms introduce some detectable bias in the sampling. We have
produced an estimate of the shape of the saddle complexity that
provides new evidence for the existence of a geometric transition in
the soft-sphere model.

Finally, our analysis of distances has shown that in this model above
$\TMC$ the system is closer to saddle points than to inherent
structures, as has been shown to be the case in some mean-field models
($p$-spin \cite{cavagna01:role}, $k$-trigonometric
\cite{zamponi03}). We have also found that there are saddles closer
than the \ac{SGM} saddle (this has been explored analytically in the
$k$-trigonometric model, where the \ac{SGM} saddle was found to be the
closest saddle \cite{zamponi03}, and the mean-field $\phi^4$ model,
where it is not \cite{andronico04}).

\acknowledgements

It is a pleasure to thank S.~Ciliberti, G.~Fabricius, J.~R.~Grigera,
and S.~A.~Grigera for discussions and suggestions. This work was
supported by Consejo Nacional de Investigaciones Cient\'\i{}ficas y
T\'ecnicas, Fundaci\'on Antorchas, and Agencia Nacional de Promoci\'on
Cient\'\i{}fica y Tecnol\'ogica (Argentina).

\bibliography{salgo}

\end{document}